\documentclass[a4paper,11pt]{article}
\usepackage[english]{babel}
\usepackage[pdftex]{graphicx}

\begin{document}

\pagenumbering{arabic}
\titlepage {
\title{The PADME experiment at LNF}
\author{Mauro Raggi$^1$\footnote{corresponding author: mauro.raggi@lnf.infn.it}, Venelin Kozhuharov$^{2}$ and P. Valente$^{3}$\\ 
{\it $^1$Laboratori Nazionali di Frascati - INFN, Frascati (Rome), Italy } \\
{\it $^2$University of Sofia ``St. Kl. Ohridski'', Sofia, Bulgaria}\\
{\it $^3$INFN Sezione di RomaI, P.le Aldo Moro, 2 - 00185 Roma, Italy}
}
\date{7/01/2015\\ to be published in the DHF2014 proceedings}
\maketitle
\abstract{
Massive photon-like particles are predicted in many extensions of the Standard Model. They have
interactions similar to the photon, are vector bosons, and can be produced together with photons. The
PADME experiment proposes a search for the dark photon ($A'$) in the $e^+e^- \to \gamma A'$ process in a positron-on-target
experiment, exploiting the positron beam of the DA$\Phi$NE linac at the Laboratori Nazionali di Frascati,
INFN. In one year of running a sensitivity in the relative interaction strength down to $10^{-6}$ is achievable,
in the mass region from 2.5 MeV $<M_{A'}<$ 22.5 MeV. The proposed experimental setup and the analysis
technique is discussed.}
\maketitle
\section{Introduction}
\label{intro}

For MeV-GeV scale dark matter the direct detection techniques are notoriously difficult. Nevertheless some of the most
appealing dark matter scenarios predict the possibility of observing new particles in such a small mass range.
This is the case of models in which the new states are hidden not because of their high mass but due
to a very small coupling to the Standard Model.
Despite attaining the highest energy ever reached at accelerators, LHC has not yet been able to provide evidence
for WIMP like particles, strongly constraining this class of dark matter models.
This largely open field of GeV-scale dark matter has recently revived models postulating the existence of a hidden sector\cite{bib:kin-mixing} interacting through a messenger
with the visible one and offers a well-motivated opportunity for experimental exploration.  Dark sector models have been used to explain
different anomalies recently observed in particles and astroparticle physics: the excess of positrons in
cosmic rays observed by PAMELA in 2008\cite{bib:pamela2008} and confirmed by AMS\cite{bib:ams-pos} and the present three sigma discrepancy
between experiment and theory in the muon anomalous magnetic moment $a_{\mu}=(g_{\mu}- 2)/2$\cite{bib:g-2-discrepancy}.  

The simplest hidden sector model just introduces one extra U(1) gauge symmetry and a corresponding 
gauge boson: the ``dark photon'' ($DP$). As in QED, this will generate interactions of the types
\begin{equation}
 \mathcal{L} ~\sim ~ g' q_f \bar{\psi}_f\gamma^{\mu}\psi_f U'_{\mu},
\label{eq:u1}
\end{equation}
where $g'$ is the universal coupling constant of the new interaction and 
$q_f$ are the corresponding charges of the interacting fermions.
Not all the Standard Model particles need to be charged under this new 
U(1) symmetry thus leading in general to a different (and sometimes vanishing) 
interaction strength for quarks and leptons. In the case of zero U(1) charge 
of the quarks \cite{bib:u1-gauge}, the new gauge boson can
 be directly 
produced in hadron collisions or meson decays. 
The coupling constant and the charges can in alternative be generated effectively 
through the so called kinetic mixing mechanism between the QED and the new 
U(1) gauge bosons \cite{bib:kin-mixing}. 
In the latter case the charges $q_f$ in equation(\ref{eq:u1}) will be just 
proportional to the electric charge and the associated mixing term in the QED Lagrangian will be 
\begin{equation}
\mathcal{L}_{mix}=-\frac{\epsilon}{2}F^{QED}_{\mu\nu}F_{dark}^{\mu\nu}.
\end{equation}
The associated mixing coupling constant, $\epsilon$, can be so small ($<10^{-3}$) 
as to preclude the discovery of the dark photon in most of the experiments carried out so far. 
Another possibility is mass mixing with the Z, 
in which case the particle could also have Z-like properties. 
If the dark photon mass is smaller than twice the muon mass and no dark sector particle lighter than the DP exist, 
it can only decay to $e^+e^-$ pairs and it is expected to be a very narrow resonance whose total decay width 
is given by:
\begin{equation}
\Gamma_{A'}=\Gamma_{A'\to e+e-}=\frac{1}{3}\alpha \epsilon ^2 M_{A'} \sqrt{1-\frac{4me^2}{M_{A'}^2}} \left( 1+ \frac{2me^2}{M_{A'}^2} \right)
\label{eq:u-width}
\end{equation}

\begin{figure}[htb]
  \begin{minipage}[t]{0.48\textwidth}
\centering
\includegraphics[width=\textwidth]{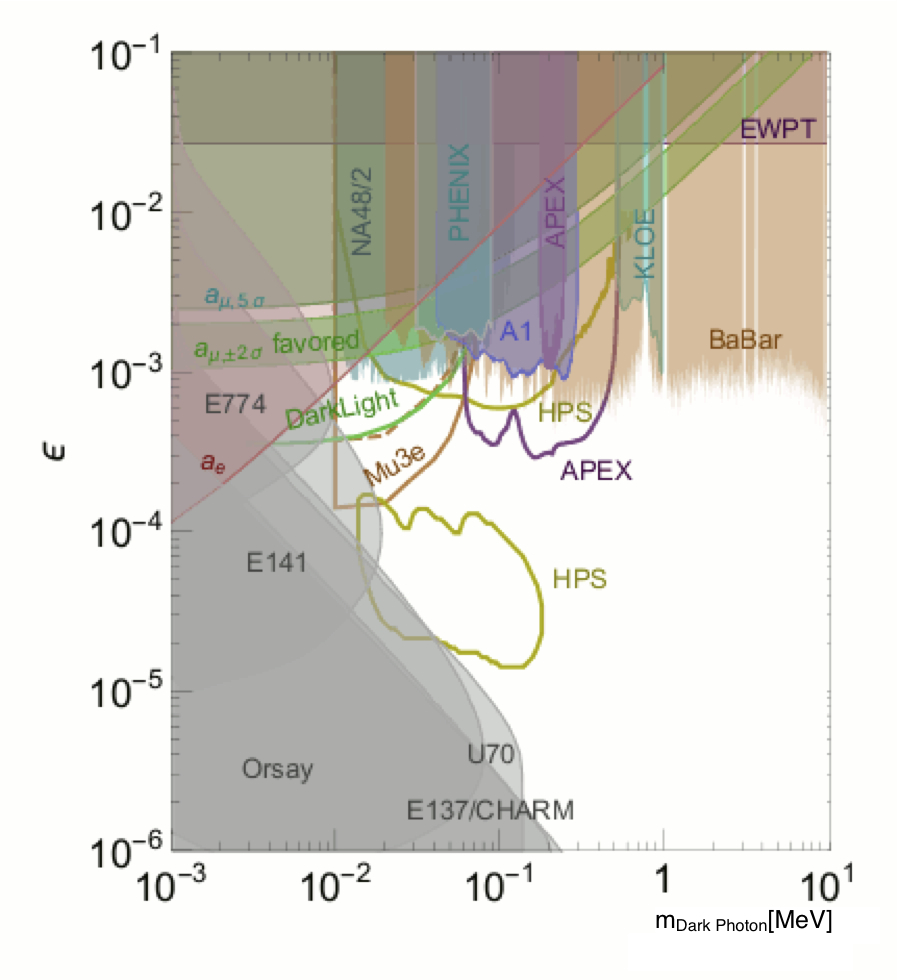}
\caption{\it Current exclusion limits and prospects for $A' \to e^+e^-$. Filled areas are excluded at 90\% C.L. Lines represent regions accessible to future experiments. Adapted from \cite{Curtin:2014cca}.}
\label{fig:StausVis}
 
  \end{minipage}\hfill
  \begin{minipage}[t]{0.48\textwidth}
\centering
    \centering\includegraphics[width=\textwidth]{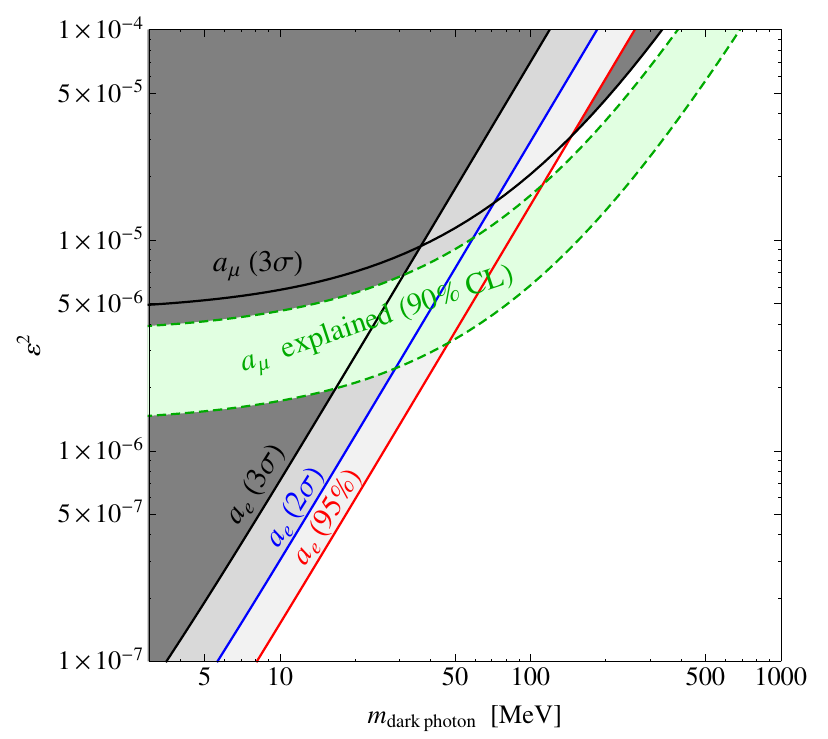}
    \caption{\it Model independent bounds for $A' \to \chi\chi$ \cite{Davoudiasl:2014kua}.}
    \label{fig:StatusInv}
  \end{minipage}
\end{figure}

which leads to a lifetime $\tau_{A'}$ proportional to $1/(\epsilon^2M_{A'})$. In this scenario many experimental constraints are available and the preferred by g-2 region has been recently ruled out by NA48/2\cite{bib:goudz}, in the hypothesis that the dark photon couples to quarks.
In Figure \ref{fig:StausVis} exclusion for $A'\to e^+e^-$ are shown.
In the most general scenario the dark sector may contain particles lighter than the dark photon itself, thus allowing the so called "invisible" decays.
The decay product $\chi$ in this case are non standard model particles which escape detection and all the decays to standard model particles are suppressed
by $\epsilon^2$ and therefore the presents exclusions are weakened. 
There are few studies on the searches of a $A'$ not decaying into Standard model particles. 
The exclusions however are in general model dependent or introduce additional parameters, namely the dark matter mass and 
coupling $M_{\chi}$ and $~\alpha_{D}$\cite{Batell:2014mga}.  
Without making further assumptions about dark sector particle masses or coupling-constants, this parameter space is only constrained by 
$(g-2)_e$, and $(g-2)_{\mu}$\cite{Izaguirre:2014bca} as shown in Figure \ref{fig:StatusInv} \cite{Davoudiasl:2014kua}.

\section{The PADME experiment}

All the dark photon searches performed so far were based on the hypothesis that the dark sector does not contain any particle of mass lower 
than that of the dark photon. After the recent improvement on the limits in the visible decay sector a new interest for invisible decays has grown in both theoretical \cite{Izaguirre:2014bca}
experimental side \cite{bib:BDX}. 
 
The Positron Annihilation into Dark Matter Experiment (PADME) aims to search for the production of a dark photon in the process 
\begin{equation}
 e^+e^- \to A' \gamma,
\end{equation}
where the positrons are the beam particles and $e^-$ are the electrons in the target.
The PADME experiment uses the 550 MeV positron beam provided by the DA$\Phi$NE linac impinging on a thin target.

\subsection{The $DA\Phi NE$ Beam Test Facility}
The $DA\Phi NE$ beam-test facility (BTF)\cite{Ghigo:2003gy}, is a
beam transfer-line from the $DA\Phi NE$ linac capable of providing up to 50 bunches per second of
electrons or positrons with 800/550 MeV maximum energy with a variable bunch width from 1.5-40 ns.
Each bunch consists of microbunches with total length of 350 ps with 140 ps flat top.
The typical emittance of the electron/positron beam is of 1(1.5)mm*mrad. 
The BTF can operate from single particle up to $10^9$ particle per bunch.
An energy upgrade of the linac is foreseen within the next three years, bringing
the maximum energy for electrons/positrons to about 1050/800 MeV together with a bunch width enlarged to ~100ns.
The sensitivity estimate for the PADME experiment assume that the $DA\Phi NE$ linac will be able
to provide 50 bunches/s of 40 ns duration with $10^4-10^5$ positrons in each. The present maximum positron energy of 550 MeV is assumed.
In these conditions a sample of $10^{13}-10^{14}$ positrons on target can be obtained in one year of data taking.

 \subsection{The PADME detector}
The PADME detector is a small scale detector composed of the following parts:
\begin{itemize}
\item {\bf Diamond Active target (50$\mu$m)}, to measure the average position, with mm resolution, and the intensity of the beam in each single bunch
\item {\bf Spectrometer}, to measure the charged particle momenta in the range 50-400 MeV
\item {\bf Dipole magnet}, to deflect the primary positron beam out of the spectrometer and calorimeter 
and to allow momentum analysis.
\item {\bf Vacuum chamber}, to minimize the unwanted interactions of primary and secondary particles.
\item {\bf Highly segmented high resolution electromagnetic calorimeter}, to measure 4-momenta and/or veto final state photons.
\end{itemize}
A schematic view of the experimental apparatus is shown in Figure \ref{fig:Exp}. Details can be found in \cite{Raggi:2014zpa}.
\begin{figure}[htb]
\centering
\includegraphics[width=12.5cm,clip]{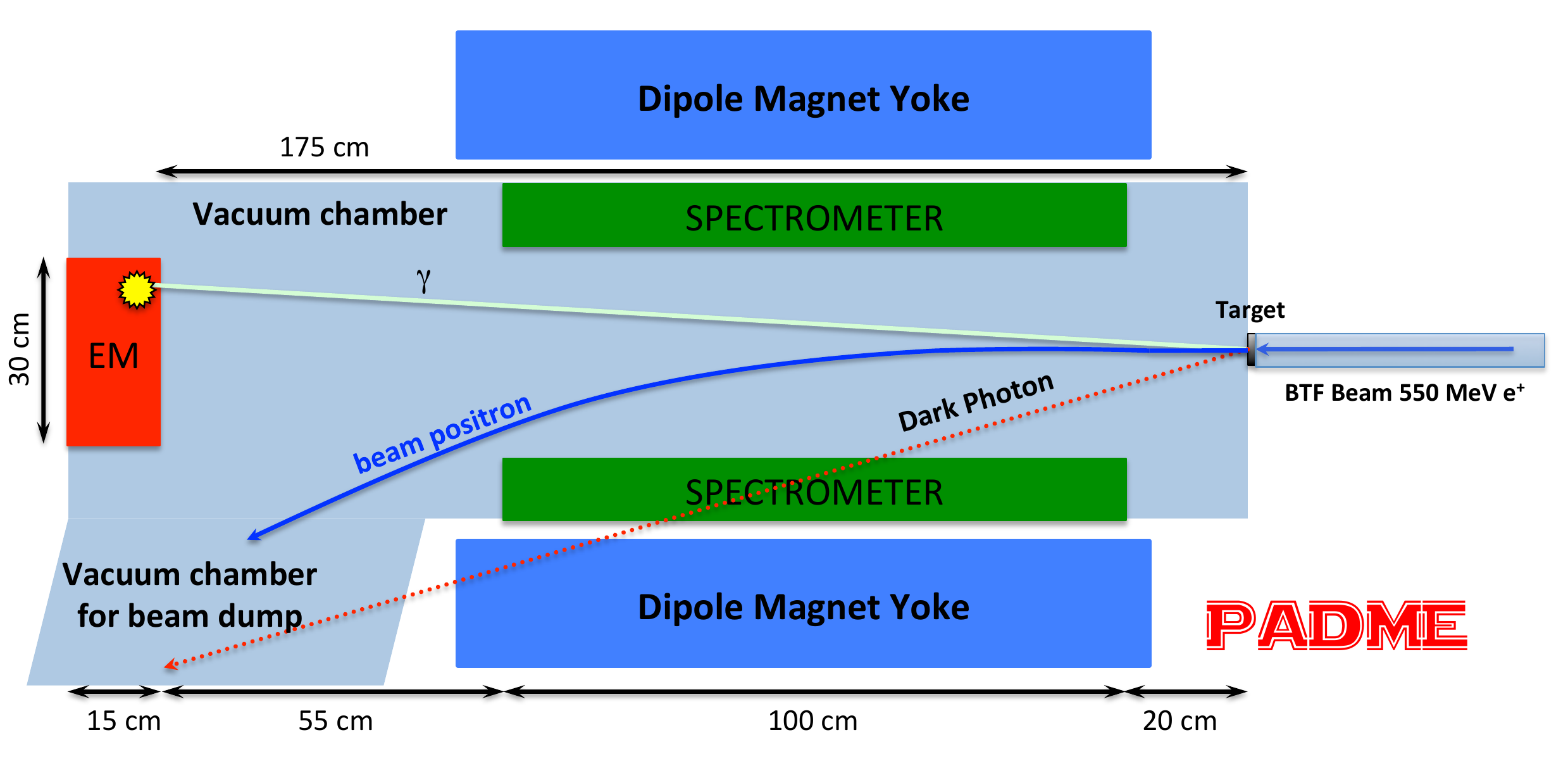}
\caption{\it Schematic of the Positron Annihilation into Dark Matter Experiment (PADME).}
\label{fig:Exp}
\end{figure}
The primary beam crosses the 50 $\mu$m diamond target and if a beam particle 
does not interact it is bent by the magnet in between the end of 
the spectrometer and the calorimeter, thus leaving the experiment undetected.
If any kind of interaction causes the positron to lose more than 50 MeV of energy the 
magnet bends it into the spectrometer acceptance, providing the veto 
capability against bremsstrahlung background.  
In case of annihilation the accompanying SM photon is detected by the electromagnetic calorimeter regardless of the $A'$ decay products. 
A single kinematic variable characterizing the process, the missing mass, is computed using the formula:
\begin{equation}
M_{miss}^2 = (P_{e^-}+P_{beam}-P_{\gamma})^2. 
\label{eq:mmiss}
\end{equation}
Its distribution should peak at $M_{A'}^2$ for $A'$ decays, at zero for the concurrent
$e^+e^- \to \gamma\gamma$ process, and should be smooth for the remaining background. 
The PADME experiment is equipped with a calorimeter and a magnetic spectrometer and it is sensitive to both visible and invisible 
dark photon decays. In fact if the $A'$ decays into $e^+e^-$, the tracks are deflected into the spectrometer and their invariant mass can be computed.

The possible $A'$ production mechanisms accessible in 
$e^+$-on-target collisions are $e^+e^- \to A' \gamma$ and $e^+N \to e^+ N A' $, 
the so called annihilation and $A'$-strahlung production.
Both process are similar to the ones for ordinary photons, and their cross section scales with $\epsilon^2$.
The PADME experiment can access four different type of dark photon searches by combining production processes
and decay final states:
\begin{itemize}
\item Annihilation produced dark photons decaying into invisible particles
\item Annihilation produced dark photons decaying into $e^+e^-$ pairs.
\item Bremsstrahlung produced dark photons decaying into invisible particles
\item Bremsstrahlung produced dark photons decaying into $e^+e^-$ pairs.
\end{itemize}
The present linac maximum positron energy of 550 MeV allows 
the production of dark photons through annihilation up to a mass of 23.7 MeV,
while masses up to 600 MeV can be reached by DP produced by $A'$-strahlung using the 800 MeV electron beams. 
At present detailed study have been performed only for annihilation production to asses
the sensitivity to invisible decays. Studies on the other final states are ongoing.
In case of annihilation production the $A'$ coupling constant $\epsilon$ can be determined by normalizing the number of
observed dark photon candidates to the Standard Model process $e^+e^- \to \gamma \gamma$ using the formula: 

\begin{equation}
  \frac{\sigma(e^+e^- \to A'\gamma)}{\sigma(e^+e^- \to \gamma\gamma)} = 
 \frac{ N( A'\gamma)}{ N(\gamma\gamma )} * \frac{Acc(\gamma\gamma)}{Acc(A'\gamma)} = \epsilon^2 * \delta , 
\label{eq:eps-calc}
\end{equation}

where $ N( A'\gamma) = N( A'\gamma)_{obs} - N( A'\gamma)_{bkg}$ 
is the number of the signal candidates after the background 
subtraction, $N(\gamma\gamma )$ is the number of observed annihilation events, 
$\delta$ is the $(e^+e^- \to A'\gamma)/(e^+e^- \to \gamma\gamma)$ cross 
section enhancement factor, 
and $Acc(\gamma\gamma) $ and $Acc(A'\gamma)$ are the 
corresponding Montecarlo acceptances for the signal and normalization channels.

\subsection{Montecarlo simulation}
To study the sensitivity of the PADME experiment to the dark photon, 
a GEANT4 simulation has been developed. 
The simulation describes in detail the segmentation
of the calorimeter and simulates showers to produces energy deposits for each
single crystal.  A cluster reconstruction algorithm providing energy and
position was implemented, starting from the energy deposits in each of the calorimeter crystals.
The magnetic field is considered to be uniform and transverse to the beam direction. The spectrometer is
modeled as an active volume from which the energy of the
crossing particles is retrieved without any reconstruction.
The simulation does not include any passive material and does not
simulate the dumping of the primary beam. To describe the bunch structure, a simultaneous multi positron
gun was implemented, taking into account beam spot size and energy spread in each single burst.
The simulation includes backgrounds simulated by GEANT4 low-energy electromagnetic libraries, two photon annihilation, ionization processes, 
Bhabha and Moller scattering, and production of $\delta$-rays. A custom generator to simulate the 
production of the dark photon and its decays into $e^+e^-$ or invisible, and the three photon annihilation
was developed.
\begin{figure}[htb]
  \begin{minipage}[t]{0.46\textwidth}
  \centering
  \includegraphics[width=\textwidth]{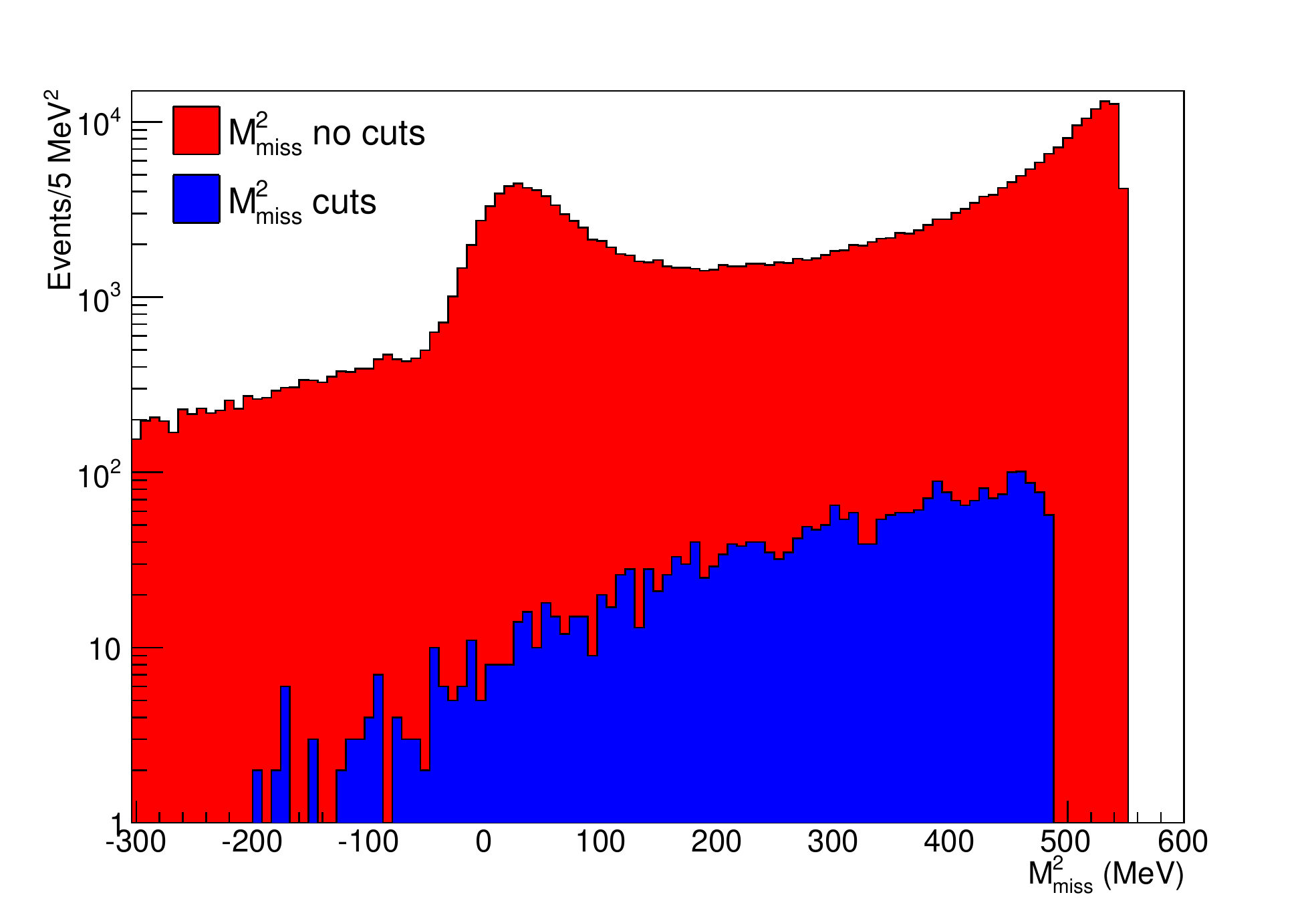}
  \caption{\it $M_{miss}^2$ for background events. In red with no selection cuts in blue after all cuts applied.}
  \label{fig:MMissBG}
  \end{minipage}\hfill
  \begin{minipage}[t]{0.46\textwidth}
  \centering
   \centering\includegraphics[width=\textwidth]{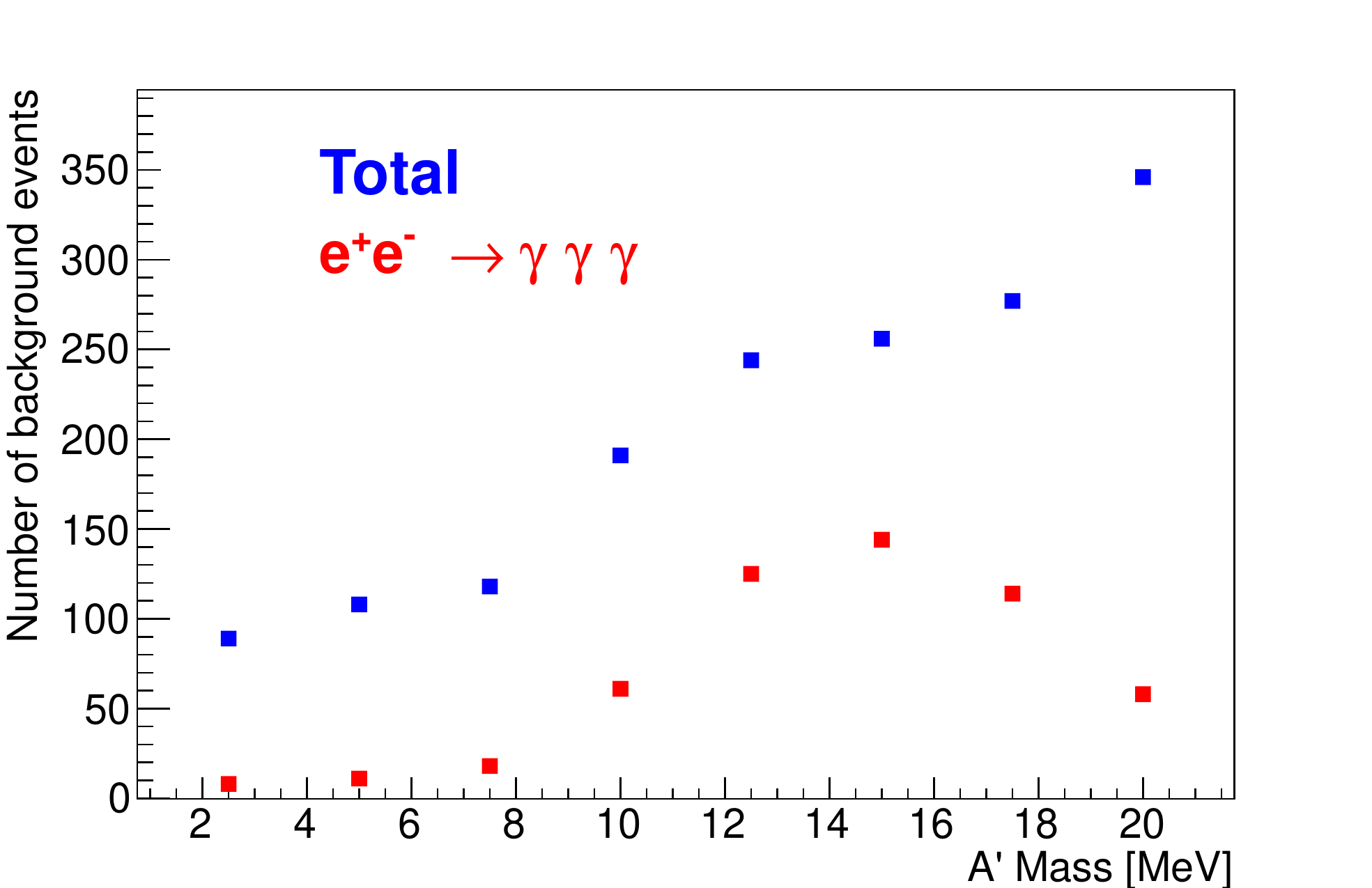}
   \caption{\it  Number of background events for different $A'$ mass hypothesis. In red $e^+e^-\to\gamma\gamma$ in blue total background after all cuts applied.}
   \label{fig:BG-dist}
  \end{minipage}
\end{figure}
A simple and preliminary selection aiming to address the possibility of 
performing a model independent search for a $A'$ has been developed. 
The selection of DP candidates is based on the missing mass squared variable, 
calculated according to formula (\ref{eq:mmiss}). 
The signal region is defined as $ \pm 1.5 \sigma_{M_{miss}^2}$ around the reconstructed value of $M_{A'}^2$,
with a proper resolution depending on the considered mass. 
It is applied both the events with visible and invisible dark photon decays. 
Since the background estimation does not depend on the $A'$ decay channel
the only difference is the acceptance for the two cases. 
With this selection, an acceptance of $\sim$20\% was achieved
for  $A'$ mass up to 20 MeV.

\begin{figure}[htb]
  \begin{minipage}[t]{0.48\textwidth}
  \centering \includegraphics[width=\textwidth]{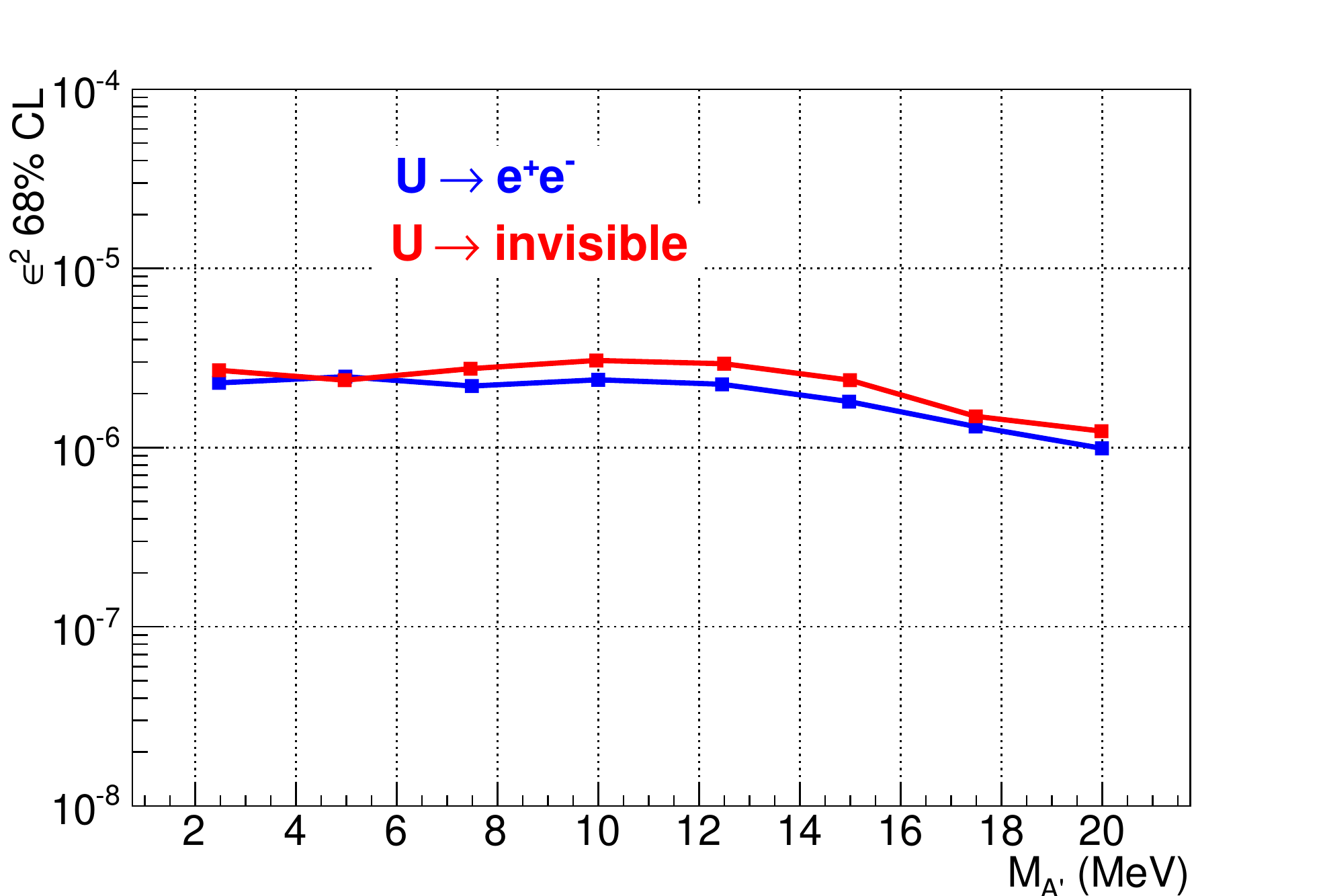}
  \caption{\it Expected exclusion limits in the $\epsilon - M_{A'}$ plane in case of no signal\cite{Raggi:2014zpa}.}
\label{fig:u-excl}
  \end{minipage}\hfill
  \begin{minipage}[t]{0.48\textwidth}
    \centering\includegraphics[width=\textwidth]{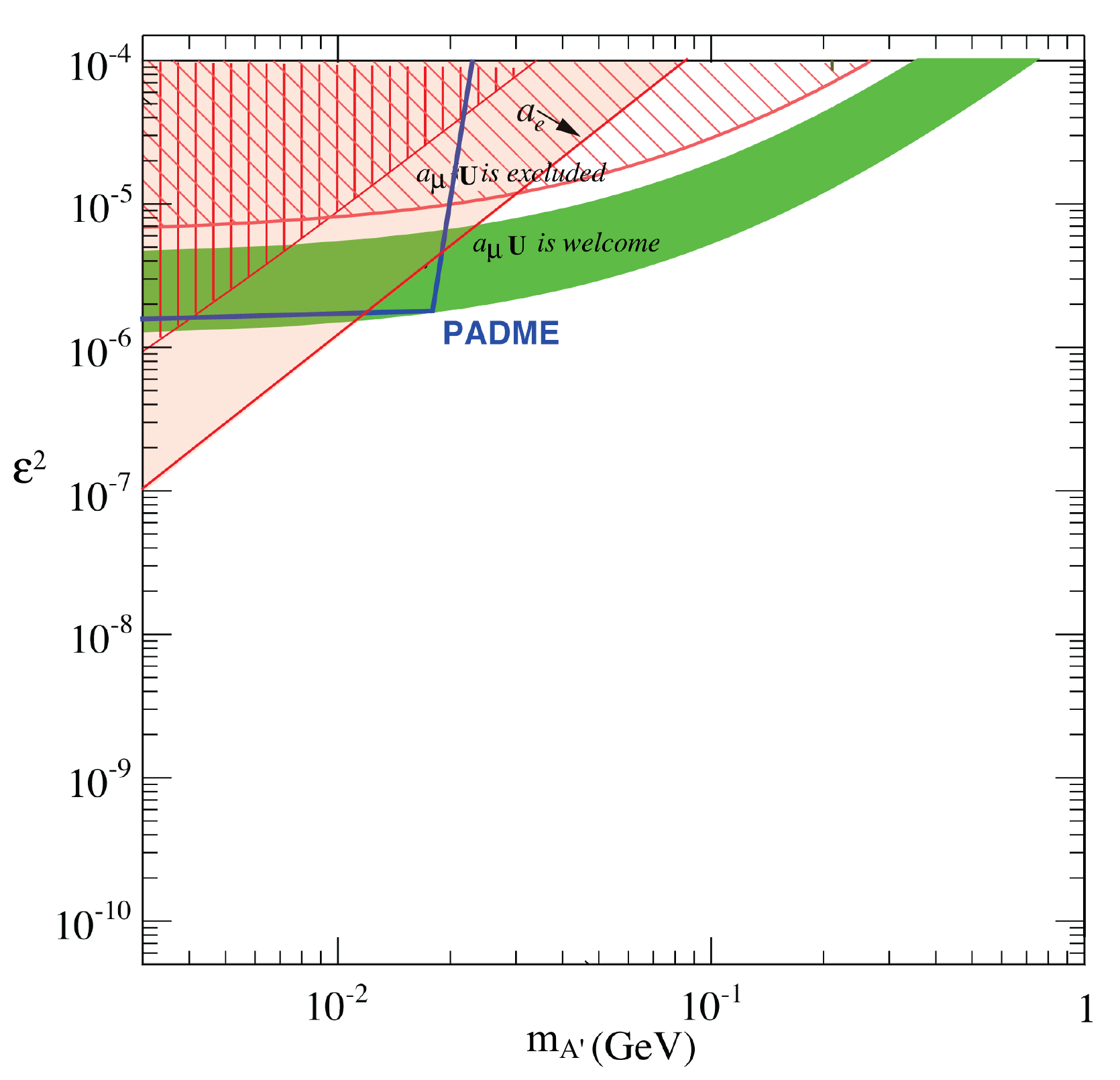}
    \caption{\it Expected exclusions in the invisible channel compared with 
  the band of values preferred by current $g_{\mu}-2$ discrepancies\cite{Raggi:2014zpa}. }
    \label{fig:u-excl-inv}
  \end{minipage}
\end{figure}

\subsection{Background estimates}

The PADME sensitivity to the invisible decay of the DP is limited by the single-photon background
originating from three main contributions: bremsstrahlung, 2$\gamma$ annihilation, and
3$\gamma$ annihilation. 
Other background processes like Bhabha scattering and
pile-up of annihilation events are included in the background
estimation through the GEANT4 simulation of the interactions
of the primary positron beam. Double annihilation
events contain extra clusters and due to energy/angle
relation are additionally suppressed with respect to single
ones. Figure \ref{fig:MMissBG} shows the $M^2_{miss}$ distribution of the simulated background
events before (red) and after (blue) the DP selection. 
The annihilation background peaks at $M^2_{miss} = 0$
the bremsstrahlung is located in the region of high $M^2_{miss}$
values while the three-photon background populates the
entire region.

\subsection{Sensitivity estimate}

With the described experimental setup and simulation technique
$10^{11}$ positrons on target were generated ($10^7$ events each with $10^4$ positron 
and $10^8$ events each with $10^3$ positron) in order to study the effect of pile up events. 
In addition, samples of 1000 events were generated for $A'$ masses 
2.5, 5, 7.5, 10, 12.5, 15, 17.5, 20 MeV, with a single $A'$. 
The background was further scaled by a constant factor of 400 
to account for one year of running of the experiment with 60\% efficiency,  
$4\cdot10^4$ positrons per bunch, corresponding to a total of $4\cdot10^{13}$ positrons on target. 
The total number of annihilation events ($N(\gamma\gamma )/Acc(\gamma\gamma)$), 
which are used for normalization, can be determined in two independent ways. 
The first is to exploit the active target for the measurement of the beam flux and  use 
the known value of $\sigma_{e^+e^-\to\gamma\gamma}$. Alternatively, direct reconstruction of the $e^+e^-\to \gamma\gamma$  
annihilation events can be performed.
Under the assumption of no signal, an upper limit on the coupling $\epsilon$ can be set, 
using the statistical uncertainty on the background as N($A' \gamma$) in equation \ref{eq:eps-calc}. 
The obtained exclusion limit, shown in Figure \ref{fig:u-excl-inv}, is very similar for visible and invisible dark photon decay 
being the selection inclusive of both cases.
PADME sensitivity to visible DP decays generated by $A'$strahlung is under investigation together the the possibility
of a dedicated beam dump experiment with $10^{20}$ EOT\cite{Raggi:WNLNF}. In this case the spectrometer is employed to reconstruct invariant
mass of the detected $e^+e^-$ pairs coming from DP decays. Mass up to $M_{A'}\sim$100 MeV can be explored with this alternative technique.

\subsection{Conclusion}
The PADME experiment aims to search for MeV scale dark matter by its decay into electrons or by detecting missing mass in low energy $e^+$ on target collision.
In particular PADME will be sensitive
to any new small mass particle produced in $e^+e^-$ collisions regardless of its decay mode and life time. Dark photons, low mass dark Higgs\cite{Batell:2009yf} or leptonic gauge bosons\cite{Lee:2014tba} models can be explored.
In addition the new-physics interpretation of any positive signal would be greatly simplified by the kinematical informations provided by the PADME detector.
Mass of the new particle and its coupling to electrons can be immediately measured. 
Early studies applied to dark photon invisible decay modes demonstrated
that PADME can reach a sensitivity down to the level of $\epsilon^2\sim1\cdot10^{-6}$ in the mass range 2.5$<M_{A'}<$22.5 MeV.
Being limited in the explorable mass region by the BTF positron beam energy, the PADME experiment will strongly profit by an energy upgrade of the existing DA$\Phi$NE linac as proposed in\cite{Valente:Eupgrade}.

\end{document}